\documentclass[10pt,a4paper,bibliography]{article} 
\usepackage{makeidx}
\usepackage{bm}
\usepackage{graphicx}
\usepackage{graphics}
\usepackage{float}
\usepackage{amsmath}
\usepackage{amssymb}
\usepackage{amsfonts}
\usepackage{latexsym}
\usepackage{epsfig}
\usepackage{eufrak}
\usepackage{euscript}            
\usepackage{dcolumn}
\usepackage{url}
\usepackage{textcomp}
\usepackage{eurosym}

\newcommand{\Acal}{{\mathcal A}}

\newcommand{\Qcal}{{\mathcal Q}}

\newcommand{\Scal}{{\mathcal S}}
\newcommand{\Ccal}{{\mathcal C}}

\newcommand{\Lcal}{{\mathcal L}}
\newcommand{\Mcal}{{\mathcal M}}

\newcommand{\half}{{\textstyle \frac{1}{2}}}

\usepackage{latexsym}
\usepackage{epsfig}
\usepackage{dcolumn}
\usepackage{pdfsync}  
\usepackage[
bookmarks,colorlinks,breaklinks]
{hyperref}
\hypersetup{linkcolor=red,citecolor=blue,filecolor=dullmagenta}

\usepackage{epstopdf}
\begin{document}

\title{The Formation of the Universe: Conjectures}

 \author{M\textsc{aurice} K\textsc{leman} }
 \maketitle
{\centering Institut de Physique du Globe de Paris 

Universit\'e Sorbonne Paris Cit\'e 

 1 rue Jussieu 75238 Paris cedex 05 France \\}

 \begin{abstract}
This article puts forward a model of the \emph{formation} of the universe, whose essential novel ingredient is a pre-Universe \textit{reservoir} $\mathrm{R_U}$ with \textit{neither space nor time dimensions}, in interaction with the universe U. U results from a process of \textit{apparition} of \textit{spacetime entities} emanating from the \textit{ur-elements} in $\mathrm{R_U}$. The analysis of this apparition relies on a few principles like the second law and the principle of conservation of energy, applied to the thermodynamic system U $\cup\ \mathrm{R_U}$, which is closed. The principle of conservation of energy does not apply to U alone. The second law must be understood as ruling the transfer of entropy from $\mathrm{R_U}$ to U, which is as small as one bit \textit{i.e.,} $k_B \ln 2$ per spacetime entity. In this context, it is shown that the pressure $p$ of the Universe is negative and that time and space are decoupled. 

We interpret the spacetime entities as elementary black holes (EBHs) at a Planckian scale. 
These EBHs with a constant entropy $k_B\ln 2$ own possibly various angular momenta and electric charges; their statistics obeys a Boltzmann distribution, if one considers the rotating and charged EBHs as high energy states of the Schwarzschild EBH. Assuming that the total mass-energy of the EBHs appeared up to the present epoch is the total mass-energy of the observable universe, $\approx 35.3\times10^{53}$ kg, we find $T\approx 4.1\ T_P$, a temperature interpreted as the temperature of apparition of the EBHs. Incidentally this model gives a possible explanation to non-local interactions, through the hidden presence of $\mathrm{R_U}$, in reason of the absence of time and space.

\end{abstract}
\newpage


\small
\newpage
\normalsize
	
$\qquad \qquad\qquad \qquad\qquad \qquad\qquad \qquad$\textit{\scriptsize Why is there \textsf{something} rather than \textsf{nothing}? For \textsf{nothing} is simpler and 
 easier than \textsf{something},} \textbf{\scriptsize Leibniz, \textsf{ Principles of Nature and Grace, Based on Reason}, 1714.}
\normalsize
\section{Introduction}
If time and space are emergent properties, they must stem from an entity where neither space nor time are present. The basic hypothesis of this work is that such an entity does exist.
Both concepts, space and time, are generally given as such; their existence is not discussed, but justified in terms of \textit{a priori} statements, \textit{i.e.} they have the status of subjective data.  This is so since Aristotle. Kant has obviously reinforced such a trend: \textit{``Space is not something objective and real, nor a substance, nor an accident, nor a relation; instead, it is subjective and ideal, and originates from the mind's nature in accord with a stable law as a scheme, as it were, for coordinating everything sensed externally."}(cited from \cite{sep2016}). For physicists (Galileo, Newton), space and time are absolute entities. For an introduction to space and time concepts through the centuries, see \cite{jammer93, reichenbach57}. Einstein posits that space and time belong to the same conceptual frame (\emph{spacetime}, since), a change of great significance, but he does not raise the issues neither of their origin nor of their ontology. In fact, as rightly claimed by Wilczek \cite{wilczek99}, Einstein's spacetime is an avatar of the Ether. Its status has not much changed with the current cosmological theories; the Big Bang is considered to occur in an already existing empty Lorentzian spacetime, the attempts to build a static, infinite, with no beginning and no end, continuously created Universe \cite{bondi48,hoyle48} have been abandoned. I do not expatiate$\dots$.\smallskip

On the other hand, a number of attempts to unify the concepts of gravitation and those of quantum mechanics have yielded a few models of the \textit{formation} of spacetime. We mention here two of them that have somewhat inspired our presentation: 
 1)- {Sorkin and coll.}, see  \textit{e.g.} \cite{sorkin03}, 
 with the building of a spacetime made of discrete elements, linked by a set of causality relations,  these events being embedded in an underlying Lorentzian manifold. 2)- {Smolin and coll.}, see \textit{e.g.} \cite{smolin2014b}, who attach a four-momentum to each  element of the foregoing kind. 

  Hereunder we assume that time and space proceed together from a mechanism of \emph{apparition} of spacetime entities $-$to be described in some detail$-$ which construct the Universe U in size, matter content, topology, from \textit{a preUniverse reservoir $\mathrm{R_U}$ devoid of time and space}; thereby $\mathrm{R_U}$ ignores the notion of dimensionality. Attempts have already been done to conceive the creation of the universe from \textit{nothing}, see \textit{e.g.} \cite{vilenkin82,verlinde11}. $\mathrm{R_U}$ could be  this nothingness, but it is endowed with specific properties and, first of all, it is a \textsf{set} of elements (in the sense of set theory), called here \textit{ur-elements}. 
    We imagine that the Big Bang is correlated to the process of apparition envisaged here. The Hubble expansion, present because of a negative pressure that comes out from the theory, ensures the growth of this Universe. The relations between $\mathrm{R_U}$ and U obey the first and the second laws of thermodynamics.
  It appears that the present model is not in conflict with the current cosmological descriptions. 
  
  Of course the ideas presented here are tentative. Cosmology is to-day in a state of affairs that allows, except for some empirical remarkable discoveries, the most speculative models. I shall not avoid speculations.

$\mathrm{R_U}$ and U possess a common thermodynamic intensive observable,
namely $\mu/T$, well defined in U where $\mu$ is the chemical potential and $T$ the temperature. Landau \& Lifshitz \cite{landau67} claim that the temperature and the chemical potential of a relativistic system in thermodynamic equilibrium are not constrained to be spatially constant but the ratio $\mu/T$ has to be. 
This intensive variable is akin in $\mathrm{R_U}$ to an algorithmic information of one bit carried by each {ur-element}.  Ur-elements are in infinite number $\aleph_0$, the first Cantor aleph. They transform at the transition $\mathrm{R_U}\rightarrow \mathrm U$ into {\textit{spacetime entities}} that constitute U, with an entropy of one bit per entity. A possible interpretation of these entities is in terms of black holes at a Planckian  scale, here called \emph{elementary black holes} EBHs.
 
 The union U $\cup\ \mathrm{R_U}$ forms a closed system, since there are no other systems. In this process time and space emerge. The transfer of elements from the reservoir $\mathrm{R_U}$ to the Universe U is irreversible. In short we suggest here that U, \textit{i.e.} the spacetime and its associated energy, is continuously formed from $\mathrm{R_U}$ ($\mathrm{R_U}\rightarrow \mathrm U$), from a process of apparition followed by a Hubble expansion.

A cosmological \textit{negative} pressure is one of the first results of our approach. Another one is that (cosmic) time and space are decoupled, as in the usual Robertson$-$Walker model of an universe obeying the cosmological principle (an homogeneous and isotropic universe). 
 If we assume that the sum of the mass-energy of the EBHs having appeared up to the present epoch is at the origin of the total mass-energy (dark energy, dark matter, ordinary matter, radiation) of the observable universe, one gets an estimation of the temperature of apparition of $T\approx 4.104\ T_P$. This is akin to a Big Bang temperature.
 
\section{The reservoir}

\subsection{structural properties, the Zermelo process}

 \normalsize Since $\mathrm{R_U}$ has no space dimensions, it is not possible to compare the ur-element sizes, which thereby are all \textit{different} from a logical point of view: the notion of congruence is meaningless in $\mathrm{R_U}$ and a spatiotemporal location cannot be used to distinguish between elements: the ur-elements, all different, form a set $\mathcal{S}$ in $\mathrm{R_U}$. According to
Zermelo, any set so defined (space and time concepts being absent) can be \textit{well-ordered} \cite{zer05} (this is also known as the \textit{axiom of choice}). We make use of this theorem: $\mathcal{S}$ is a \textsf{set} that can be well-ordered.  
  In other words it makes sense to choose a first ur-element, then a second, and so on, and in this way to order all the ur-elements starting from a first one.  
 This structural property is usually indicated as follows:
\begin{equation}
\label{1}
a\prec b\prec c\prec\dots
\end{equation}
where $a$, first chosen, precedes $b$, which precedes $c$, and so on. That process is set in `outside' $\mathrm {R_U}$, and this sequential apparition builds U: we call it a \textsf{Zermelo process}.
 We argue that it gives some validity to the concept of time, 
see also \cite{smolin2014b}; the interval in the sequence of choices can be interpreted as a unit time lapse. Thus \textsf{time separates from space}. This is reminiscent of the \textsf{cosmological principle}, according to which 
time and space are decoupled.
 
 The question of the embedding of the sequence of spacetime entities into a Lorentzian manifold has been much discussed.  More precisely, Sorkin and coll. \cite{sorkin03,dowker2013} have considered a partially ordered set of such entities, a \emph{poset}, called a \emph{causet} if given a causal structure, (the class of posets includes obviously the subclass of linear order sets under consideration in Eq.~\ref{1}) with a Poisson distribution embedded into a Lorentzian manifold. We shall hypothesize that the Zermelo process $\mathrm{R_U \rightarrow U}$ builds such a discrete set of entities with its Lorentzian embedding. The fundamental property of a causet is indeed that it implies the Lorentz group \cite{malament77}. The essential addition we make to the theories of Sorkin {and coll.} and of Smolin {and coll.} is an analysis of the thermodynamic properties of this process. Their use of a partially ordered set instead of a linearly ordered set is valid equally well in the framework of our model, yielding subsets of simultaneously appearing spacetime elements, but there is no necessity for this complication in the frame of the operation we describe.

\subsection{algorithmic entropy of the ur-elements}

One cannot attach an entropy to the set $\Scal$ of ur-elements in $\mathrm{R_U}$ since there are no space coordinates neither time: the notion of randomness is absent, the entropy of disorder is vanishing. On the other hand, because of the multiplicity of orderings induced in the Zermelo process $\mathrm{R_U}\rightarrow\mathrm{U}$, each Eq.~\ref{1} ordering can be thought of as a microstate of $\Scal$.

Assume that there is an enumerable infinite number $|\Scal| =\aleph_0$ of ur-elements, where $|\Scal|$ denotes the cardinal of $\Scal$. The cardinal of the class $\mathcal T$ of distinct order types of enumerable sets (\cite{kamke50} p. 67, theorem 2),  is $|\mathcal T|=2^{\aleph_0}$.  
 The entropy of $\Scal$ can then be \emph{formally} written
$$\Sigma_\mathrm{R} = k_B\ln|\mathcal T|=k_B \aleph_0\ln2,$$
and interpreted as the sum of the (all equal) entropies attached individually to each ur-element in $R_U$, namely \begin{equation}
\label{2}\sigma = k_B\ln2 .\end{equation} Thus $\sigma$ has the character of 
 an \textit{algorithmic entropy} \cite{chaitin76,hutter2007}  attached to a unique entity. 
 Our derivation of this result does not claim to be rigorous, but yields results of a convincing nature, as it will appear.
Eq.~\ref{2} satisfies the rule of additivity of the entropy: the entropy of two non-intersecting subsets of respectively N$_1$ and N$_2$ ur-elements $\in  \mathrm{R_U}$ is the sum N$_1k_B\ln2  + \mathrm N_2k_B\ln2 $ of the entropy of each subset; it also tells us that \textsf{the missing information carried by each ur-element is exactly one bit}, a most reasonable and fascinating result. Eventually we surmise that $\sigma$ enters the entropy balance of U, identifying the missing information to a physical entropy in U \cite{landauer96}.

Zermelo's theorem applies equally well to a continuous set. We shall not investigate this possibility and keep to the assumption that $\Scal$ is enumerable. This has some interesting consequences, as we see later, end of para.~\ref{inf}.

\section{From the reservoir to the universe} \label{transition} 

\subsection {what is borrowed from classical physics}\label{borrow}
We cannot start from scratch in the present theoretical development of the formation of the Universe, and we have to rely on a certain number of laws or axioms, although one might expect that this number is small and that the theory we develop should itself shed light on some of the laws of physics we apply daily.\smallskip

The\textit{ cosmological principle} is deduced from the causal structure, as discussed above. We also hypothesize that 
\\$-$ the first and second laws apply to the evolution of the closed system $\mathrm{R_U} \cup \mathrm{U}$, 
\\ $-$ the two subsystems $\mathrm{R_U}$ and U can be given intensive and extensive thermodynamic variables with usual definitions and properties. This requires some care for $\mathrm{R_U}$. In these conditions, it will be shown that, without further assumptions, the entities cannot return from U to $\mathrm{R_U}$, as if a \textit{semipermeable} membrane were separating them. Such a situation leads to identify the energy density, in fact its opposite $- \varepsilon$, to an \textit{osmotic pressure}, even if a semipermeable membrane between U and $\mathrm{R_U}$ is a figment of our imagination,
\\ $-$  the Zermelo process is attended by the appearance of energy and entropy in U. This entropy was above qualified of algorithmic, as if the transition from U to $\mathrm{R_U}$ could be described as resulting from a Turing machine process (it cannot be given an \textit{a priori} explanation in statistical mechanics terms), since it is believed that any physical process can be simulated by an universal computing device.
 
  In conclusion this paper is built on some concepts of classical thermodynamics and general relativity, in which framework it appears a certain number of non trivial features, which we develop now.

\subsection {{thermodynamic potentials in $\mathrm{R_U}\cup \mathrm U$.}}\label{eqstate}
There is no point in distinguishing different thermodynamical potentials in $\mathrm{R_U}$; volume is not a relevant concept and the temperature $T$ is not fixed because of the vanishing of the entropy $S_\mathrm{R}$. The only quantity we are allowed to refer to is the energy variation $\delta E_\mathrm{R}$ in a Zermelo process; we can write:
\begin{equation}
\delta G_\mathrm{R}=\delta F_\mathrm{R}=\delta W_\mathrm{R}\qquad=\delta E_\mathrm{R}.\label{3} \end{equation}
Observe that the Gibbs energy $G_\mathrm{R}$, the free energy $F_\mathrm{R}$, the enthalpy $W_\mathrm{R}$, the energy $E_\mathrm{R}$ are undefined thermodynamic functions, because not only they are infinite, but also there are no finite densities attached to them; the concept of density is ignored in $\mathrm{R_U}$. The same remark holds for the number of ur-elements $N_\mathrm{R}=\aleph_0$. Therefore, we shall not use these notations, which are meaningless. On the other hand $\delta G_{\mathrm{R}}$, $\delta F_{\mathrm{R}}$, $\delta W_{\mathrm{R}}$, $\delta E_{R}$, $\delta N_{\mathrm R}$ have a physical meaning, inasmuch as the total system $\Sigma_{tot}=\mathrm{R_U} \cup \mathrm{U}$ is closed. This yields:
$$\qquad  \delta E_\mathrm{R}+\delta E=0, \ \ \textrm{{energy conservation}}; \qquad  \delta N_\mathrm{R}+\delta N=0, \ \ \textrm{logical relation};$$ where $\delta N=\sum_i \delta N_i$, a summation over the different entities of constant $\sigma=k_B\ln 2$ generated at the transition.

$N=\sum_i N_i,\ \,{E=\sum_i  N_i \varepsilon_i v_i=N\varepsilon v},\ \, V=N v,\ \, T,\,\dots$ are well defined thermodynamic functions, living in U, $\varepsilon$ is a positive quantity, which measures the mean energy density transferred from $\mathrm{R_U}$ to U in a process of apparition, $v$ is the mean volume attached to one entity. They depend on the age of the Universe. 
 
 Hence we have, assuming that there is no exchange of heat and that the total system $\Sigma_{tot}=\mathrm{R_U}\ \cup$~U is closed:
\begin{equation}
dE=-dE_\mathrm{R}= \quad\varepsilon v\, dN +N\, d(\varepsilon \, v) \equiv \varepsilon\, dV+V\, d\varepsilon, \label{4}\end{equation} where we have applied the first principle to the total system $\mathrm{R_U}\ \cup$ U.  A more detailed account of the thermodynamics implied by Eq.~\ref{4} is given later; but as a first approximation Eq.~\ref{4} suggests to interpret $\varepsilon$ as a \textit{negative pressure}; 
\begin{equation}
\label{5}
p=-\varepsilon.
\end{equation} This equation of state is well-established in cosmology and is considered at the origin of the inflation of the universe
\cite{mukhanov2012} at its very early beginning. It is strictly valid at the first appearing spacetime entity, when $V=0$.

The foregoing considerations do not provide the $\mathrm{R_U}\, \rightarrow \,$U Zermelo process with special features and are not very appealing, {except for the existence of a {negative pressure}}. Generally, the thermodynamic equilibrium between two systems \textit{in contact} implies the equality of the temperatures $T_a=T_b$ and of the chemical potentials $\mu_a=\mu_b$. The assumption that $\mathrm{R_U}$ and U are in equilibrium is a notion difficult to assess insofar as there is no geometrical contact (but in a sense $\mathrm{R_U}$ and U are in contact everywhere where U exists) and furthermore there is no defined temperature in $\mathrm{R_U}$. As stated above, the chemical potential and the temperature of a relativistic system in thermodynamic equilibrium are not constrained to be spatially constant; but the ratio $\mu /T$ has to be. This latter remark suggests a solution to this puzzle; we shall assume that the quantity $\mu /T$ relative to the spacetime entities is not only constant through a given spatial Universe at a given cosmic time, but is constant in time, by reason of the status of $\mathrm{R_U}$. This constancy expresses the equilibrium of U with the reservoir $\mathrm{R_U}$. 
 
 \subsection{the entropy of a spacetime entity}\label{entropy}

The variation of the internal energy of the Universe with time contains two contributions, one from 
the already $N$ existing spacetime elements, \emph{i.e.} the Hubble's expansion, one from the appearance of $d N$ new entities emerging from the reservoir, see Eq.~\ref{4} . All together one has:
\begin{equation}
\label{6}
dE\,(= TdS -pdV+\mu dN)=d(N\varepsilon v).
\end{equation}

Introducing the mean entropy per object $s,\ S=Ns$, this equation also reads:
\begin{equation}
\label{7}
(\mu+sT)dN+ N(Tds-pdv)-pvdN -d(N\varepsilon v) =0. 
\end{equation}

This expression can be simplified by employing the Gibbs-Duhem relation $d\mu = -sdT+vdp$ (where $d\mu$ is the same for all the spacetime entities), thereby writing $(\mu+{Ts})dN = dN(\mu+{Ts})-Nvdp-NTds.$ One gets eventually:\begin{equation}
\label{8}
d\big\{N[(\mu+{Ts})-Nv(p+\varepsilon)]\big\} =0.
\end{equation}

By integrating, and assuming that the constant of integration vanishes, 
one gets: 
\begin{equation}
\label{9}
s = -\frac{\mu}{T} +\frac{v(p+\varepsilon)}{T}.
\end{equation} 
Eq.~\ref{9} also obtains, with likewise no constant of integration, in a classical, statistical, derivation of the thermodynamical properties of an expanding universe, cf. \cite{mukhanov2012}, p. 82. 

In Eq.~\ref{9}, the chemical potential term $-{\mu}/{T}$ is an intrinsic contribution of any spacetime entity, related to the nature of this entity; the second term relates to the embedding of this entity in U. Thus 
\begin{equation}
\label{10}
 \frac{\mu}{T}=-k_B\ln 2 \end{equation} $T$ being usually positive, the chemical potential of the spacetime entities is negative.
 
\section{Cosmological results}\label{cos} 

 \subsection{elementary black holes as possible spacetime entities}\label{inf} 
 
   We assume that the entities that appear are \emph{elementary black holes} each with an entropy of one bit: the area of the event horizon $\Acal = 4\Scal=4\ln 2$ is a constant but these emerging entities might differ by their charge $\Qcal$ and their angular momentum $\Lcal$, thereby by their mass $\Mcal_{\Lcal,\Qcal}$.
 
  Let us first consider a black hole with no charge and no angular momentum (a so-called Schwarzschild black hole): the horizon area $\Acal$ and the mass $\Mcal$ are given by the Bekenstein$-$Hawking formulae \cite{penrose2016} p. 270; in Planck units: 
 \begin{equation}
\label{11}
\Acal=4 \ln 2 = 2.772,\ \quad\Mcal_{0,0}=
\half \sqrt{\frac{\ln 2}{\pi}} = 0.2349.
\end{equation}  
 These are the smallest black holes possible in the framework of a  classical description (noted EBH, elementary black hole, in the sequel). As stated by Hawking \cite{hawk71}:  \textit{`since gravitational collapse is essentially a classical process, it is probable that black holes could not form with radii less than the Planck length'} precisely these EBHs. \textit{$\dots$ One might therefore expect collapsed objects to exist with masses from $10^{-5}\ \mathrm g$ upwards.' } 
 
 In the general case $\Lcal\neq 0$, $\Qcal \neq 0$, we have the relation \cite{christodoulou71} :
\begin{equation}
\label{12}
\Mcal^2_{\Lcal,\Qcal} =\frac{\Acal} {16\pi} +\frac{4\pi \Lcal^2}{\Acal}+\frac{\Qcal^2}{2}+\frac{\pi \Qcal^4}{\Acal},
\end{equation} with the condition \begin{equation}
\label{13}
\Lcal^2/\Mcal^2_{\Lcal,\Qcal} +\Qcal^2\leqslant \Mcal^2_{\Lcal,\Qcal}.
\end{equation}
 \indent At constant $\Acal$, \textit{i.e.}, at constant entropy, $\Mcal_{\Lcal,\Qcal}$ is always larger than $\Mcal_{0,0}$. 
 
\emph{More on condition Eq.~\ref{13}}. The equality is obtained for $$\Mcal^{2}_{\pm} = \half (\Qcal^2 \pm\sqrt{\Qcal^4 +4\Lcal^2})$$. The condition \ref{13} reads $\Mcal^2 \geqslant
 \Mcal^{2}_+$, that can also be written:
 \begin{equation}
\label{14}
\Mcal^2_{\Lcal,\Qcal}\geqslant \frac{\sqrt{\Qcal^4 +4\Lcal^2}}{2}+\frac{\Qcal^2}{2} \ \Rightarrow \ 
\Big({\sqrt{\Qcal^4 +4\Lcal^2}}-\frac{\Acal} {4\pi}\Big)^2\geqslant 0,
\end{equation} which is satisfied for any value of $\Qcal$ and $\Lcal$. The equality sign corresponds to the so-called `extremal' black holes.\smallskip
 
\textit{A remark about the magnitude of the entropy.}
 A EBH of $10^{-5}$ g and one bit is equivalent in mass to a population of $\approx 1.3\times10^{19}$ protons, thereby an extraordinarily small entropy for a very large number of particles. There are approximately $10^{80}$ baryons in the observable universe. Their total entropy, spread all over the sequence of Eq.~\ref{1}, would then be approx. $10^{61}$ bits, 
 a definitively small quantity compared to what can be expected at the present epoch by Penrose \cite{penrose04} , at least $10^{21}$ bits per baryon according to this author.  \smallskip
 
 We shall assume that the EBHs resulting from the Zermelo process thermalize, forming a bath of `particles' labeled $\ell,q$ ($\Lcal= \ell$, $\Qcal=q$). These particles will be thought of as occupying different energy levels $\Mcal_{\Lcal,\Qcal}$   \begin{equation}
\label{15}
\Mcal^2_{\ell,q} =\Big(\frac{\ln 2} {4\pi} + \ell^2\frac{\pi }{\ln 2}+\frac{1}{2} q^2+\frac{\pi}{4\ln 2}q^4\Big)=\frac{m^2_{\ell,q} }{m_P^2},
\end{equation} where $m_{\ell,q}=\Mcal_{\Lcal,\Qcal} \ m_P$ ($m_P$ Planck mass).  All these entities are discernible, due to their origin in successive processes of apparition. This suggests that altogether they obey a Boltzmann statistics and are distributed proportionally to:
\begin{equation}
\label{16}
n_{\ell,q}={g_{\ell,q}}\ \exp{-\frac{m_{\ell,q}}{k_B T}}={g_{\ell,q}}\ \exp{-x\Mcal_{\ell,q}},\ \  \mathrm{where} \ \
x=\frac{m_P c^2}{k_B T}=\frac{T_P}{T},
\end{equation} ${g_{\ell,q}}$ being the number of internal degrees of freedom.

 EBHs can be interpreted as boson `particles'. The ur-elements in R$_\mathrm U$ cannot be given individual descriptions although they are all different from the point of view of set theory, they can transform into any $\Mcal_{\Lcal,\Qcal}$ in U, each $\Mcal_{\Lcal,\Qcal}$ able to be reproduced many times in U. Furthermore the chemical potential $\frac{\mu}{T}=-k_B \ln 2$ is negative, Eq. \ref{10}, a property that characterizes bosons.
 
 The labels $\ell$ and $q$ must be discrete. If ever $\ell$ and $q$ were continuous, then the set $\Scal\in \mathrm R_\mathrm U$ would also be continuous; more generally  the set $\Scal$ and that one whose elements are labelled by \{$\ell,\ q$\} have the same cardinality. 
 Thus $\Lcal$ and $\Qcal$ are quantized. We choose the usual rules of quantization: $e \Qcal= \pm e q, \ \hbar \Lcal = \hbar\ell; \  \ell, q\in \mathbb{Z}$; $2\ell+1$ microstates associated to $\ell$ and two opposite electric charges associated to $q^2$. Therefore
$$q= 0 \rightarrow g_{\ell,0} = 2\ell+1,  \qquad  q\neq 0 \rightarrow g_{\ell,q} = 2(2\ell+1).$$
\subsection{data} \label{num} 

In this section, we recall some observational data, taken from ref. \cite{wiki2019a} and \cite{bennett2013} (WMAP), which we shall compare in the next section \ref{iso} with numerical values calculated for the distribution above, Eq.~\ref{16}.

 \subsubsection{observational data.}
 \noindent $-$ age of the observable universe $$t_0 =13.799 \times 10^{9}\ \mathrm{years}=4.355\times 10^{17} \ \mathrm{sec},$$  
$-$ radius
$$r_{0}=4.4 \times 10^{26}\ \mathrm m,\ \rightarrow \frac{4}{3}\pi r_0^3=3.57\times 10^{80}\ \mathrm m ^3$$ $-$  the total mass-energy density being $9.9\times10^{-27}$ kg/m$^3$ and the volume of the observable universe $3.57\times 10^{80}$ m$^3$, the total mass-energy (including ordinary matter, radiation, dark matter and dark energy) is 
$$m_{tot}= 35.3\times 10^{53}\ \mathrm{kg}; $$

\subsubsection{mass of the observable universe.}
\emph{Planck values} related to the present theory. Let us consider U at two instants $t$ and $t+t'_P$, $t'_P$ being the lapse of time for the apparition of one spacetime entity of size $\ell'_P$, mass $m'_P= \Mcal_{0,0} m_P$ (Eq.~\ref{11}). We define $\ell'_P =ct'_P$ by the relation $\Acal \ell_P=4\pi \ell'_P{^2} $; thus:\\
$-$ Planck values related to our theory: $$t'_P={0.4697} t_P, \quad \ell'_P=  {0.4697}\ell_P, \quad m'_P =0.2349m_P .$$ 
 $-$ number of transfers from R$_U$ to U that build the observable universe $$n'_0=t_0/t'_P=1.72\times 10^{61}.$$

   If we assume that all the EBHs are Schwarzschild's black holes, with no angular momentum and no charge, their contribution to {the total mass-energy} of the universe would be 
$$m'_{0}=  \Mcal_{0,0}\ m_P \  n'_0 =m'_P\  n'_0 = 8.79 \times 10^{52}\ \mathrm{ kg}.$$
 This value falls in a reasonable range of values, which fact justifies somewhat the approach we have taken. It is nevertheless much smaller than $m_{tot} $ above and compares better with $m_{0}=1.62 \times 10^{53}\ \mathrm{kg}$, the total mass of ordinary matter (4.6 \% of the total mass-energy). But the present theory does not introduce any other source of matter-energy than the EBHs, whose decay should therefore be associated with the appearance of dark matter, dark energy, radiation, etc.  Thus one expects that the mass of the EBHs should rather compare with $m_{tot} $. To make this possible, it is necessary to introduce in the energy balance the other EBHs of the same entropy $\Mcal_{\ell,q}$, $\ell, q\neq 0$. In fact, there is no reason to exclude them of our account.
 
 \subsection{isentropic EBHs as bosonic particles; mass-energy}\label{iso}
 
\begin{figure}
\includegraphics[width=4in]{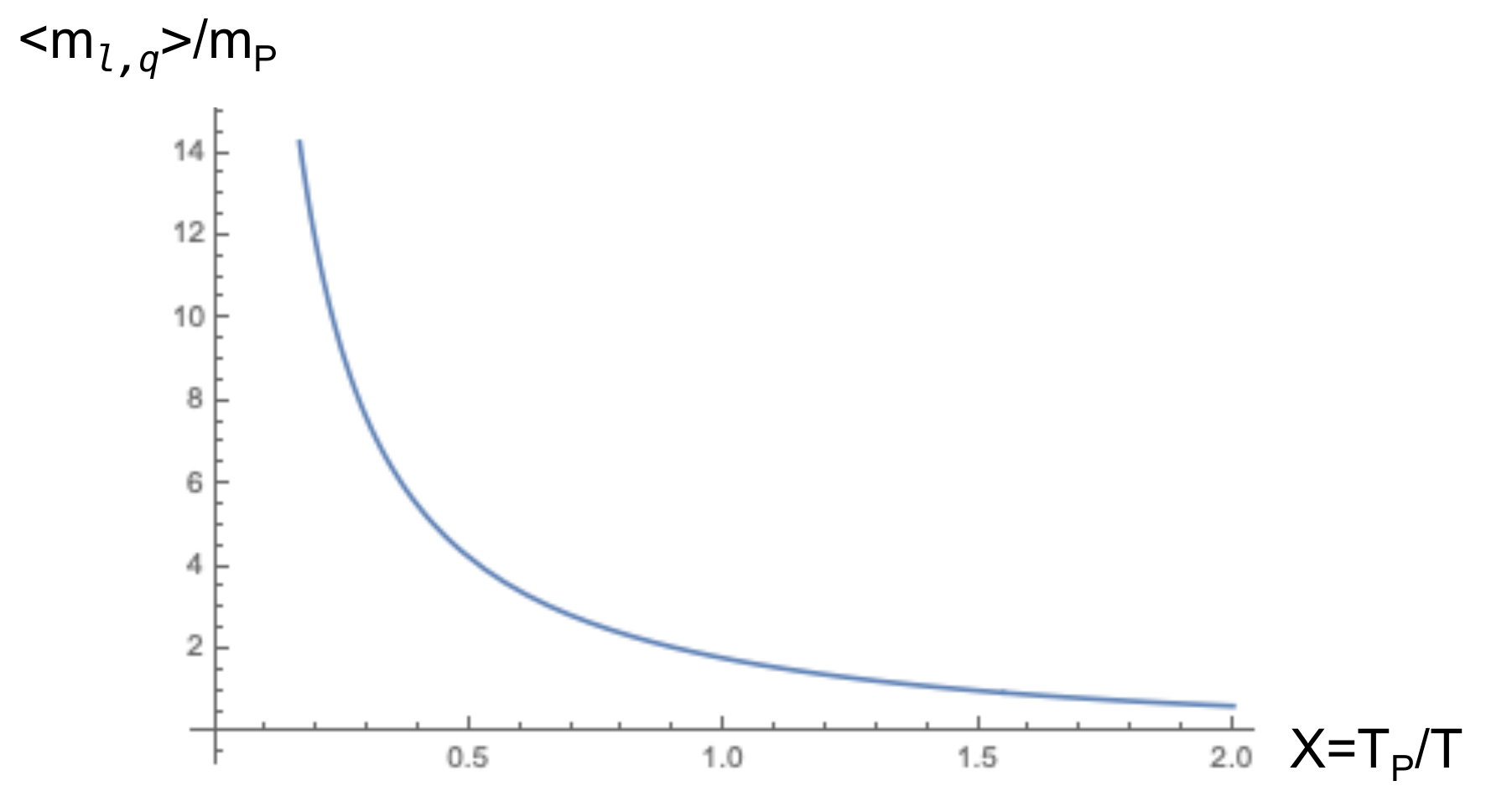}
\caption{\footnotesize Relation between the mean mass value $<m_{\ell,q}>$ of the  EBHs and their temperature of apparition T.}
\label{f2}
\end{figure}

    According to their distribution Eq.~\ref{16}, the mean mass per EBH at a temperature $T$ can be written: 
 \begin{equation}
\label{17}
<m_{\ell,q}>=\frac{E}{Z},\qquad\mathrm{where:}\ \ E=\Sigma_{\ell,q}m_{\ell,q} \ n_{\ell,q}, \quad Z=\Sigma_{\ell,q} \  n_{\ell,q}.
\end{equation}
\begin{figure}
\begin{center}
\includegraphics[width=5in]{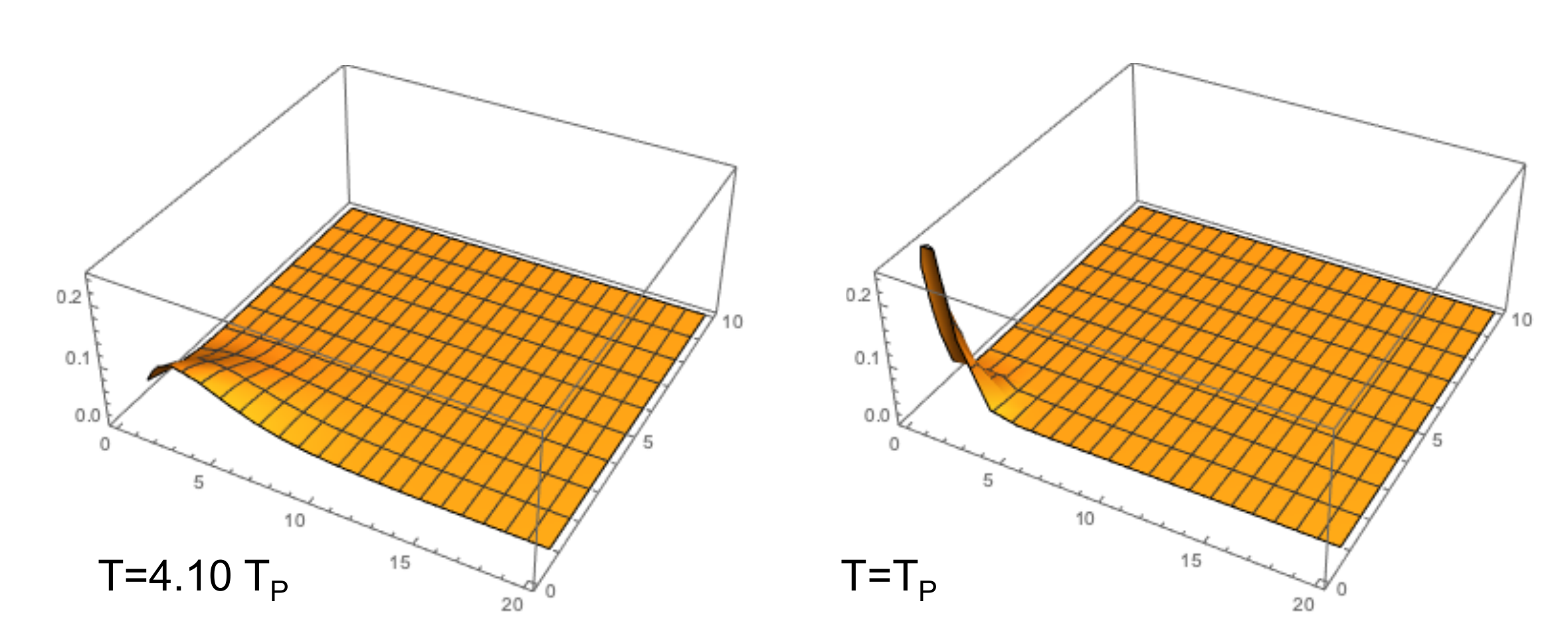}
\caption{\footnotesize 
Probability $p_{\ell,q}= n_{\ell,q}/Z$ for $x = 1$ (right) and $x = 0.24365$ (left); $0\leqslant\ell\leqslant20,\ 0 \leqslant q\leqslant10$. The intersections of the mesh correspond to integral values of $\ell$ and $q$.}
\label{f1}
\end{center}
\end{figure}The double summation should in principle be taken on the $n'_0$ EBHs,  a stupendous number. In fact this summation converges relatively fast, and a numerical calculation reveals that $E$ and $Z $ do not differ significantly whether the double summation be taken in the range $0<\ell, q<50$ or $0<\ell, q<1000$ (see Fig.~\ref{f1}, where it is visible that the probabilities $p_{\ell,q}=n_{\ell,q}/Z$ are rather flat in the range $\ell, q>10$). We assume that the total mass-energy of these EBHs is at the origin of the total mass-energy of the cosmos $m_{tot} = 35.3 \times 10^{53}$ kg, as calculated above; therefore the mean mass $<m_{\ell,q}>$ of a EBH should be $<m_{\ell,q}>=m_0/n'_0=9.427 \ m_P$. Taking $E/Z=9.427 \ m_P$ one gets $x=0.24365$, see Fig.~\ref{f2}. 
 To summarize:
\begin{equation}
\label{18}
<m_{\ell,q}>=9.427\ m_P, \qquad T = 4.104\ T_P.
\end{equation}
\indent Fig.~\ref{f1} (probability distributions of ($\ell,q$) EBHs for $T=T_P$ and $T=4.10 T_P$, the value we retain) and Fig.~\ref{f2} ($<m_{\ell,q}>$ in function of the temperature) summarize the results. Notice that the probabilities, as expected, are the greater for the smaller $\ell,q$ values, but that the distributions are very sensitive to the temperature.  It appears in particular that most of the EBHs have small $\ell,q$ values when $T=T_P$, whereas one expects a more scattered distribution for $T=4.10\ T_P$. The question arises whether with this value of T we are still in a Planckian range, or sub-Planckian.

 Of course, the type of calculation made here assumes implicitly that the EBH ensemble is at thermodynamic equilibrium, which requires that the time of decay by a Hawking evaporation process  $t_{ev}$ of these black holes is long enough to allow for the apparition of a macroscopic number of EBHs. The existence and stability of EBHs has been much debated recently, the question being whether or not they disappear by the process of Hawking evaporation. It has been advanced that evaporation could cease at Planck scales, yielding `remnants', in reason of a generalized uncertainty principle \cite{adler2001} or the effects of extra-dimensions and quantum-gravitational spacetime fluctuations \cite{carr2005}. The so-called `extremal' black holes, which obey Eq.~\ref{13} and Eq.~\ref{14} with the equality sign, do not Hawking-evaporate and are consequently stable \cite{kallosh92}. It has also been suggested that there may be `relics' formed \textit{before} the Big Bang, in a cyclic universe \cite{brandenberger2017,rovelli2018a}. This latter suggestion is in a sense not in contradiction with our model. Finally it has been advanced that they constitute a large part of dark matter \cite{macgibbon87, rovelli2018b}. According to ref.  \cite{adler2001}, the characteristic time of the Hawking evaporation of a micro black hole (of Planckian size) is of order  $t_{ch}=4.8 \times 10^4\ t_P$. This seems to be large enough to justify the calculated distribution, inasmuch as a small number of `particles' suffices for a fast convergence of the summations Eq.~\ref{12}, as already stated. 
  
  \textit{A remark about extremal black holes}.
  Eq.~\ref{14}, with the sign of equality, reads: 
  \begin{equation}
\label{19}
q^4 + 4\ell^2=\Acal^2/16\pi^2 .
\end{equation}
 Eq.~\ref{19} is represented by a circle $\Ccal$ centered at the origin in the plane [$2 \ell,\ q^2$]. Thus $|\ell|\leqslant \Acal/8\pi =0.1103$, $|q|\leqslant \half\sqrt{\Acal/\pi}=0.4697$, and $\ell$ and $q$ do not take together integer values along $\Ccal$. 
  $\Mcal_{\ell,q}$ (Eq.~\ref{15}), denoted $\Mcal_{q}$ hereunder, reads:
\begin{equation}
\label{20}
\Mcal_{q}=\sqrt{\frac{\Acal} {8\pi} +\frac{1}{2} q^2},
\end{equation}  
whose maximum $\Mcal_{q_{max}} = 0.4697$ is small compared to the expected mean value $<m_{q}> =<\Mcal_{q}> m_P$  of Eq.~\ref{18}, viz. $<m_{\ell,q}>=9.427\ m_P$. Thus extremal black holes do not appear in the Zermelo process.
 
 \section{Some final remarks}\label{diss}

There is of course something much `unphysical' in  R$_\mathrm U$, which appears \textit{a priori} as non-falsifiable, non testable, as much weird as a multiverse. Its introduction in physics requires the use of a profound theorem of the mathematical set theory, $-$this part of mathematics finds here some application to physics for the first time, whereas there is seemingly no other domain of mathematics that could not find already some application. But the consideration of R$_\mathrm U$ has this advantage over the multiverse to predict a negative pressure $p=-\varepsilon$ Eq.~\ref{3}, a small entropy (one bit per entity), a decoupling of (cosmic) time and space, and to yield a reasonable Big Bang temperature, taken the magnitude of the mass-energy of the observable universe at the present epoch. The only point where it differs from the standard theory is that the Big Bang is in the present theory an everlasting process, with however an origin, the apparition of the first spacetime entity in the Zermelo sequence Eq.~\ref{1}, $-$thus the name of Big Bang is not overindulged. This difference is worth a future investigation. Notice that the present results belong to a trans-Planckian domain, but obtain from very simple thermodynamical arguments. A complete theory of quantum gravity implying space and time as emergent entities is still missing, but see \cite{damour2008}. It would perhaps give a more physical significance to R$_\mathrm U$, which might be nothing more than a \textit{representation} at the classical level of a sub-Planckian object.

A remark about the one-bit entropy per EBH; it would mean, if interpreted in terms of statistical physics, that there are only two microstates pertaining to a EBH. Which microstates? The present favored interpretation is in terms of quantum entanglement, as first proposed by Bombelli \textit{et al.} \cite{bombelli86}, for a review see \cite{bekenstein94}. This entropy would then measure the quantum correlations between the exterior and the interior states. It is intriguing to remark that all those states, whatever their distance in U, whatever indeed the non-locality of their interactions, are in contact with $\mathrm{R_U}$, where the concepts of distance and time are absent. This specific nature of non-locality in our model can well be related to the non-local effects observed experimentally, see \textit{e.g.} \cite{aspect82}. The same remarks should apply to any kind of spacetime entity whose entropy stems in quantum entanglement. 
  
  Notice finally that there are no antiparticles in this model; $\mu$ does not change sign (antiparticles and particles have opposite chemical potentials). Therefore antiparticles appear only during the decay or evaporation process of black holes.
  
However, for negative temperatures, the chemical potential would be positive. This possibility remains to be explored. 
  
\section*{Acknowledgments}
  This study contributed to the IdEx Universit\'e de Paris ANR-18-IDEX-0001. I am grateful to Prof. T. Damour, Prof. J.-P. Poirier and Prof. M. Veyssi\'e for fruitful discussions.

\bibliographystyle{unsrt}
\bibliography{biblioarc1}
\end{document}